# Affective Factors in STEM Learning and Scientific Inquiry: Assessment of Cognitive Conflict and Anxiety

Lei Bao[1], Yeounsoo Kim[1], Amy Raplinger[1], Jing Han[1], Kathleen Koenig[2]

[1]*Department of Physics, The Ohio State University*
[2]*Physics and STEM Education, University of Cincinnati*

*Abstract*

Cognitive conflict is well recognized as an important factor in conceptual change and is widely used in developing inquiry-based curricula. However, cognitive conflict can also contribute to student anxiety during learning, which can have both positive and negative impacts on students' motivation and learning achievement. Therefore, instructors need to be informed of the impacts of introducing cognitive conflicts during teaching. To get this information, teachers need a practical instrument that can help them identify the existence and features of cognitive conflict introduced by the instruction and the resulting anxiety. Based on the literature on studies of cognitive conflict and student anxiety, a quantitative instrument, the In-class Conflict and Anxiety Recognition Evaluation (iCARE), was developed and used to monitor the status of students' cognitive conflict and anxiety in the *Physics by Inquiry* (PBI) classes. This paper introduces this instrument and discusses the types of information that can be measured. Research and pedagogical values of this instrument are also discussed.

# INTRODUCTION

## Theoretical Framework on Conceptual Change

The study of conceptual change in students has been a major area of research in science education on constructivist learning for more than three decades (e.g., Duit & Treagust, 2003; Treagust & Duit, 2008). Many constructivists in science education have argued that cognitive conflict is an important factor in conceptual change and a useful strategy in inquiry based education (Chan, Burtis, & Bereiter, 1997; Chinn & Brewer, 1998; Dreyfus, Jungwirth, & Eliovitch, 1990; Kang, Scharmann, & Noh, 2004; Kim & Kwon, 2004; Kwon, Lee, Park, Kim, & Lee, 2000; Kwon, Park, Kim, Lee, & Lee, 2000; Lee, Kwon, Park, Kim, Kwon, & Park, 2003; Limón, 2001; Posner, Strike, Hewson, & Gertzog, 1982; Strike & Posner, 1992).

In the process of learning through science inquiry, students often come into the classroom with established beliefs embedded within a wide range of diverse everyday contexts. Many of such beliefs are non-scientific with some being strongly held and difficult to change. Therefore, helping students to "change" their non-scientific preconceptions to the expert beliefs has been the main goal of many of the studies on conceptual change. Through research, it has been found that by explicitly recognizing the discrepancy between their current beliefs and the scientific ones (often referred to as the experience of a cognitive conflict), students can be motivated to change their current beliefs, which starts the processes of conceptual change. A favorable conceptual change also depends on how students resolve the cognitive conflict and manage a number of affective issues such as interest and anxiety that are induced by experiencing cognitive conflict.

Posner et al. (1982) identified four requirements for successful conceptual change. Students must have (1) dissatisfaction with their current conceptions, and they must see the new conception as (2) intelligible, (3) plausible, and (4) fruitful. Simply put, students need to first recognize that there is a conflict between their current views and the new information to be learned, and if they are going to reject their old views, the new idea needs to make sense to them. Posner et al. also noted that students must take the conflict seriously. This opens the door to many affective factors, such as interest, motivation, and anxiety, which can impact the process of conceptual change.

Dreyfus et al. (1990) conducted interviews with ninth- and tenth-grade biology students to determine what issues can arise during cognitive conflict. They found that the type of knowledge under consideration affects how well it is accommodated. Topics which they describe as "experience bound" (those that students experience in everyday life) lead to meaningful conflict, but there is much less impact when students are faced with conflict in the "cultural knowledge" domain (topics learned only in school). Dreyfus et al. also found that even if a student appears to experience conceptual change, the outcome may not always be the targeted expert views. The student can develop new but still incorrect conceptions. This implies that instructors need to be aware of this possibility and provide students an opportunity to express new conceptions.

Another important consideration in conceptual change is student ability. Dreyfus et al. (1990) and Zohar and Aharon-Kravetsky (2005) found that students with a high level of ability benefit from cognitive conflict while lower ability students either fail to appreciate the cognitive conflict or do not get as much out of it. According to Dreyfus et al., bright, high achieving students enjoy cognitive conflict and the "flabbergasting effect" brought on by the conflict and its resolution. Lower achieving students do not share this attitude. Cognitive conflict is something that such students would try to avoid, and the feeling of anxiety is threatening. Rather than embrace a conflict and attempt to resolve it, lower achieving students chalk it up as a failure on their part. The differences between how high and low achieving students handle cognitive conflict accounts for much of the mixed outcomes of research done on the effectiveness of cognitive conflict. Looking at a class as a whole, there can be a cancelling effect from the two groups. It is important, then, for instructors to know the ability levels of their students before using cognitive conflict as a teaching strategy.

Chan et al. (1997) conducted a study where students were presented with ideas in a way that either minimized or maximized cognitive conflict. They found that situations in which conflict was maximized led to higher-level "knowledge building," in their words, and more successful conceptual change. This has important implications for curriculum development and classroom structure. Chan et al. also found that success was affected by age and whether students worked alone or in groups. With older students (grade 12), group work led to higher-level knowledge building activities while younger students (grade 9) seemed to benefit more from individual work. Chan et al. stated that this was not always the case and the results could be highly context dependent, as different pairs of students might handle a conflict in different ways.

In their study, Lee et al. (2003) developed a cognitive conflict process model to explain cognitive conflict in terms of four constructs: recognition of an anomalous situation, interest, anxiety, and cognitive reappraisal of the conflict situation. This model connects cognitive conflicts with anxiety linking the cognitive side of learning with the affective factors. Lee et al. also found that anxiety has both positive and negative effects on student learning. If a student feels anxiety in the form of frustration or fear, he/she will not be motivated to learn the material, and the cognitive conflict is actually destructive to learning; but if the student feels anxiety in the form of interest or a need to resolve the cognitive conflict, he/she may be motivated to pursue further learning, and the anxiety will have lead to a positive outcome.

The works by Dreyfus et al. (1990), Zohar and Aharon-Kravetsky (2005), and Lee et al. (2003) point to a consensus view that cognitive conflicts and the associated anxiety can lead to mixed impacts on students' learning behaviors, depending on education settings and student backgrounds. The results are also contrasted by several other studies on feelings of anxiety during academic situations, which reported a negative correlation between school achievement and anxiety measures (Cassady & Johnson, 2002; Hembree, 1988; Hong & Karstensson, 2002). The contextually dependent mixed impact of anxiety on learning raise a strong call for more careful studies in how anxiety may be induced in learning contexts and how such anxiety may impact students of different backgrounds. The development of iCARE is an attempt to address this need by providing a valid and easy to use instrument for measuring and monitoring students' cognitive conflicts and the induced anxiety in changing learning contexts.

## Present Work on Science Inquiry and Cognitive Conflicts

During learning, cognitive conflicts are inevitable (Hong & Karstensson, 2002). This is especially so in courses that emphasize scientific inquiry such as *Physics by Inquiry* (PBI) (McDermott et al., 1996).  PBI is an inquiry-based learning environment where students work in groups of three or four in a laboratory setting to discover, with guidance, various physics concepts. During the class time, students are actively engaged in a series of interactive inquiry learning processes receiving guidance and immediate feedback, contributing ideas and new thoughts, and evaluating evidence and hypotheses through discussions with peers and instructors. Figure 1 shows the typical inquiry learning cycles in a PBI classroom.

Through in-depth study of simple physical systems and their interactions, students gain direct experience with the process of science inquiry. Starting from their own observations and predictions, students develop basic physical concepts, use and interpret different forms of scientific representations, and construct explanatory models with predictive capability. The primary emphasis is on discovering through guided investigations, dialogues between the instructor and individual students, and small group discussions. A major goal is to help students think of physics not as an established body of knowledge but rather as an active process of inquiry in which they can participate and in which teaching is done by questioning rather than by telling.

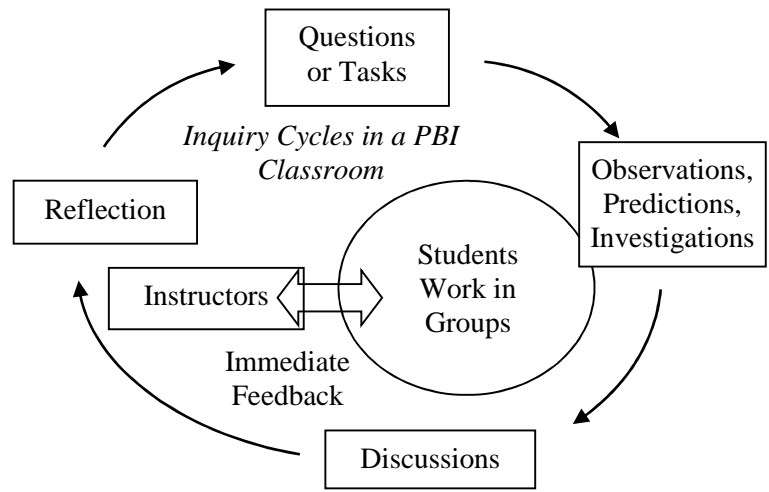

Figure 1. Inquiry cycles in a typical PBI classroom and in PBI homework.

With the goal of helping students develop reasoning skills to conduct effective science inquiry, PBI has been widely used for preparing preservice and in-service K-12 teachers to teach physics as a process of inquiry (McDermott, Shaffer, & Constantinou, 2000). The PBI methodology has also been adopted in many science courses to help underprepared students succeed in the mainstream science courses that are the gateways to majors in science, technology, engineering, and mathematics (STEM) (Redish, 2003).

The students who take PBI are of university age, and many of them are science education majors (preservice teachers). In the class, students typically work in groups of 3 or 4 following the PBI text, which guides them through new concepts, having them perform experiments and answer questions. At various points throughout a lesson, the text has students stop for a checkpoint; at this time, the students discuss what they have learned with an instructor (professor or expert-level teaching assistant) who asks them thought provoking questions, induces group discussions, and challenges them with alternative hypotheses, experiments, and transfer problems, etc. until all students in the group demonstrate a solid understanding of the underlying concepts. Class runs for 2 hours and 48 minutes with two classes each week. Grades in PBI are broadly distributed over a range of formative and summative assessment components including class participation, checkpoints, homework, journal reports on learning experiences, questions of the day, and exams (two midterms, one final). Usually, the exam scores count towards about 40% of the total grade, which can help reduce students' test-related stress and anxiety and allow students to focus on each individual step of the entire learning process.

For the PBI class, the most emphasized teaching strategy is to have students construct knowledge from seeing and resolving conflicts among peer students, between students and instructors, and between a student's present understanding and new information. Therefore, instructors and researchers need to know when and under what circumstances students may experience cognitive conflict and the extent to which students are impacted affectively (in terms of anxiety) by the conflict. Knowing such information can greatly improve the instructors' capability in guiding students to properly address their anxiety. Such anxiety can have both positive and negative impacts to student learning; addressing it properly means that it can become a productive force for learning. It is then important for researchers and teachers to be equipped with practical tools that can conveniently measure cognitive conflict and the associated anxiety during learning.

However, there is limited availability of instruments in this area. The primary instrument available presently is the Cognitive Conflict Levels Test (CCLT), developed by Lee et al. (2003), which measures information about the existence and degree of students' cognitive conflict during learning. The CCLT is a 12-item 5-point Likert scale test that includes four measurement components of cognitive conflict: (a) recognition of an anomalous situation (the existence of conflict), (b) interest, (c) anxiety, and (d) cognitive reappraisal of the situation.

Although useful, CCLT does have limitations. For example, the structure of CCLT is designed to measure the cognitive conflict of a single observation event; therefore, it cannot be used to measure the multiple occurrences of cognitive conflict in a practical teaching environment, which is a very dynamic process that typically involves multiple complex situations that could trigger cognitive conflict in different students. In addition, CCLT measures students' recognition and the resulting feelings of the cognitive conflict triggered by a given condition; however, it does not measure the information on what contexts trigger the conflict and how they do so, which is crucial for understanding and improving student learning.

Inspired by CCLT, we have developed a new survey instrument, the In-class Conflict and Anxiety Recognition Evaluation (iCARE), which can be implemented conveniently in classrooms to measure the status of students' cognitive conflicts and their levels of anxiety resulted from experiencing conflicts. The design of this instrument targets two main assessment constructs: (1) the context elements of the instruction and curriculum that trigger cognitive conflicts and (2) students' feelings and reactions in responding to their cognitive conflicts. In the following parts of the paper, we discuss the development of this instrument and its implementation in a PBI course on electric circuits.

# METHOD: The Design of iCARE

## Context of the Development

There have been many studies in the literature about cognitive conflict, conceptual change, and the affective aspects of learning such as anxiety (Cho & Kim, 2004; Kim, Acar, & Bao, 2005; Lee et al., 2003; Limón, 2001; Shin, Kwon, & Kim, 2005; Treagust & Duit, 2008). The results of these studies provided the theoretical and experimental basis that shaped the design of iCARE. In particular, the development of iCARE builds heavily on the Cognitive Conflict Levels Test (CCLT) (Lee et al., 2003), which was designed to measure students' experience of cognitive conflict caused by observing a demonstration. Each CCLT item reflects a particular aspect of cognitive conflict. The procedures for implementing CCLT were that students first took a pretest about their preconceptions on a scientific demonstration problem. Students were then shown a demonstration, after which a posttest on their beliefs about the demonstration was given along with the CCLT test. On the CCLT test, students indicated how much each item reflected their feelings using a 5-point Likert scale. Based on research, CCLT has been shown to be a valid and reliable testing instrument (Lee et al., 2003).

There are several limitations to CCLT. First, while the CCLT can be used in many contexts, its format limits its use to immediately after a demonstration. Since a class session usually contains multiple instruction units, it is then practically difficult to use CCLT in a classroom setting to obtain measures of conflict and anxiety which may have risen from different phases and components of a complete class session. Second, although CCLT measures the level of students' cognitive conflicts and anxiety, it doesn't provide any information on how the conflicts and anxiety are caused in real education contexts, which is of great importance for educators and researchers in understanding students' learning difficulties and finding ways to help students improve their learning. Third, CCLT does not provide an open-ended section for additional discussions that may occur during a class; if students have feelings that cannot be adequately expressed with the items in the CCLT, then researchers do not have an opportunity to indentify these feelings without conducting interviews. Finally, CCLT was designed and tested for demonstration-based instruction. This leaves the realms of inquiry-based learning and group learning completely unexplored. For these reasons, CCLT is a good tool for use in controlled research settings in which researchers can target a limited number of pre-determined situations, but it is hard for instructors to use CCLT in practical teaching scenarios. In fact, CCLT seems to have been only used in research settings and there hasn't been any report on using CCLT in teaching practices.

In responding to the limitations of CCLT, the design of iCARE makes an emphasis on measuring several key features that are missing in CCLT. iCARE explicitly measures the context situations that lead to the cognitive conflicts and anxiety. The situations are modeled after the features of inquiry-based group learning environment for seamless alignment with the PBI course. For example, cognitive conflict and anxiety can be triggered by discrepancies between ideas from different students in a group conducing group discussions and investigations. This information is explicitly measured in iCARE to allow teachers and instructors to pinpoint the contexts in which the conflict and anxiety originate. The format of measurement in iCARE is also flexible, making the instrument easily adaptable into courses using different teaching settings, such as lectures, recitations, and labs.

The design of iCARE also makes an emphasis on the important role that teachers can play in delivering effective inquiry based instruction that utilizes cognitive conflict. Since inquiry based teaching methods are becoming more prominent in current and future curriculum, teachers need to understand and be empowered to deal with the multi-faceted aspects of learning in science inquiry.

In reality, however, teachers tend to have limited views of teaching, which could be due to the fact that teachers are often excluded from the education research process and therefore are not as informed as they should be (Treagust & Duit, 2008). If the goal is to have students benefit from constructive use of cognitive conflict, teachers must have an understanding of how cognitive conflict affects learning and in what contexts it is likely to occur. This brings about the need to inform teachers about the cognitive underpinning and education values of conflicts and anxiety and the need to provide them the necessary instrument to formatively assess and address cognitive conflicts in teaching practices. iCARE is designed to bridge the gap between teachers and researchers and can be used as a formative assessment tool in professional development programs for training pre-service and in-service teachers to deliver effective inquiry based learning. As teachers are trained in these techniques, iCARE can be used to help teachers learn what to look for in their students and how to help students of different backgrounds and at different stages of learning.

In summary, iCARE addresses the issues in CCLT and puts emphasis on the practicality of using the instrument in real teaching. iCARE can be completed in just a few minutes at the end of each class period (or after completion of a learning section), measures multiple situations that can cause cognitive conflicts, gives a quantitative scale of the level of the resulting anxiety, and includes an open-ended section should students want to elaborate on their feelings. The design of iCARE makes it possible for teachers to assess students' cognitive conflict and anxiety in everyday teaching on a regular basis.

## Measurement Components

As discussed earlier, Lee et al. (2003) targeted four aspects of cognitive conflict in CCLT—recognition of an anomalous situation, interest, anxiety, and cognitive reappraisal. iCARE integrates the CCLT measures to produce four components that explicitly measure the context that triggers conflict and the students' reactions to the conflict: (a) situations that lead to discrepancies in learning which can further trigger cognitive conflicts, (b) student recognition of experiencing cognitive conflict, (c) an estimate of anxiety level, and (d) student reactions in responding to conflict situations. This measurement structure follows closely the central theme of cognitive conflict, which suggests that the discrepancies in learning can cause the learner to experience cognitive conflict and learning-related anxiety. The learner's level of anxiety and reactions to conflict are dependent on the background of the learner, which in combination will affect the post-conflict learning trajectories with both positive and negative pathways possible (Dreyfus, Jungwirth, & Eliovitch, 1990; Pintrich, Marx, & Boyle, 1993; Lee et al. 2003; Zohar and Aharon-Kravetsky, 2005).

The first part of iCARE is designed to identify specific contexts in instruction that lead to cognitive conflict among students. In a group-learning, inquiry-based course such as PBI, cognitive conflict can be triggered by a number of situations leading to discrepancies between (1) a student's expectations and observations of outcomes of demonstrations or experiments, (2) a student's multiple alternative conceptions (including the new concept being introduced), (3) a student's understanding and a peer's understanding, and (4) a student's understanding and the information delivered by the instructor.

The identification of these different situations provides important information about the settings of the instruction that may have contributed to students' experiencing cognitive conflicts.  In addition, studies have shown that different situations can have different impacts on conceptual change and cognitive development (Hashweh, 1986; Druyan, 2001; Piaget, 1950).  For example, when peer conflicts arise in a group learning environment, students need to come up with a solution that satisfies not just an individual student but the entire group, which usually requires more effort and puts more pressure on students as they go through multiple cycles of group discussion and negotiation about evidence, opinions, and hypotheses. The affective and educational impacts of conflict within a group can certainly be different from teacher-student conflict, where the teacher is often regarded as the authoritative figure who usually directs the learning trajectories in a set of predetermined pathways (Druyan, 2001). Information as to what situations lead to cognitive conflict is important to researchers and teachers and needs to be measured and carefully considered in curriculum development and teaching.

The second part of iCARE is designed to identify the feelings that students experience during a cognitive conflict. Based on CCLT, we chose to probe three types of feelings that are considered typical in a cognitive conflict (Lee et al., 2003). These are (A) "The differences surprised me," (B) "The differences increased my interest in the topic," and (C) "The differences made me want to pay more attention to the topic and spend more time working on it." These feelings are closely associated with the recognition of cognitive conflict and may affect how students take the next step to either resolve or ignore the conflict.

The third part of iCARE provides an estimate of the level of anxiety experienced by students during a cognitive conflict. Researchers have suggested two components of anxiety: a cognitive component and an emotional component (Druyan, 2001). Additional sources that may lead to anxiety, such as test anxiety, are also well studied (Ball, 1995; Cassady and Johnson, 2002), but in this study, we focus on the cognitive component of anxiety, which is believed to be more directly related to learning and task performance (Hembree, 1988). The design of iCARE includes three items addressing cognitive based anxiety that are directly related to students' experiencing cognitive conflict. These items were based on CCLT (Lee et al., 2003) and modified to include (A) "The result of this experiment confused me," (B) "Since I can't resolve the differences, I am uncomfortable," and (C) "I am upset because I cannot understand the reason for the result." To complete the measurement, students are asked to evaluate each item using a 5-point Likert scale (1 = "not at all true", 5 = "very true"). Then students are asked to calculate the total rating score of the three. If the total score is less than 9, the student is considered to have a low level of anxiety; if the score is 9 or above, the student is considered to have a high level of anxiety (Lee et al., 2003; Kwon et al., 2000a; Kwon et al., 2000b). Based on this calculated score, students are guided to complete one of the two groups of items in part four of iCARE.

The fourth part of iCARE identifies students' reactions and behaviors in responding to conflict situations. Students choose a response from one of two groups of items according to their anxiety scores. These items were selected based on previous studies of students' anxiety-related behaviors in cognitive conflict (Johnson & Johnson, 1995; Kim, 2002; Cho & Kim, 2004). In order to identify additional types of anxiety-related behaviors, the fourth part of iCARE has an open-ended item for students to report cases that are not included. All eight items are summarized in Table 1 along with the types of behaviors that researchers used to categorize the items in the literature. Students are allowed to select only one option; while this does limit the richness of feedback received, it also helps identify the best candidates for descriptions of student reactions to cognitive conflict.

The four sections of iCARE follow the flow of a progressive activation of meta-cognitive self-reflections of one's learning. Therefore, iCARE can also function as an epistemic training tool. For example, the first part of iCARE explicitly prompts students to think about the processes and specific experiences of learning (rather than the content), which is not a common practice among students. Students may not initially put much thought into what has happened in a class period, but iCARE promotes self-reflection on the learning process. When iCARE is used regularly, students can get in the habit of being aware of their learning experiences. In addition, iCARE repeatedly prompts students to recognize the process of knowledge development through experiencing and resolving cognitive conflicts; therefore, it can help students realize the constructive nature of knowledge and that encountering conflicts is common in the process of developing scientific understanding. Gradually, these self-reflections can guide students toward developing a more favorable metacognitive and epistemic standing.

Table 1. Students' reactions and behaviors in responding to cognitive conflict

| Anxiety Level | Types | Items on iCARE |
|---|---|---|
| Low | Agreed predictions | "Before the experiment, I predicted multiple possible outcomes. From the experiment, I have seen one of my predictions proved. So I am satisfied with the experiment result even without detailed explanations." |
| | Confidence in resolving conflict | "I was confident that by reevaluating my previous beliefs, I would be able to find an explanation without others' help." |
| | Dependence on others' ideas | "I accepted what instructors or my classmates had said. I didn't spend much effort to find an explanation on my own." |
| | Use of past personal experience | "I made my predictions for this experiment by thinking about my past experience. I also tried to make sense of what I saw in the experiment based on my understanding through that experience." |
| High | Confidence in preconceptions and inconsistency in understanding | "Before the experiment, I was highly confident in my previous understanding of the subject. However, my understanding seems to be inconsistent with the outcome of the experiment." |
| | Re-inspection of reasoning for predictions | "After I saw the outcome of the experiment, I tried to explain it by considering things that I might have ignored as I was making the predictions." |
| | Lack of self-confidence | "I believe that there must be good reasons that can explain the experiment well. But right now I don't think I have learned enough physics to build a good explanation yet." |
| | Recognition of inability to resolve conflict | "In this experiment, the results are inconsistent with what I expected based on my experience and I haven't been able to resolve the problem yet." |

A common concern on a survey instrument like iCARE is the fidelity of students' self-reports. It can be argued that students may not have enough metacognitive understanding when reporting their cognitive conflict and anxiety. The design of iCARE addresses this issue by using mostly "descriptive" items relating to students' actual experiences of conflict situations (contexts) and feelings. The anxiety measures are embedded within these descriptions, which can be extracted by researchers. Students do not explicitly evaluate their anxiety and are not aware that they are being measured on this. The words of "anxiety" and "conflict" never show up in the iCARE instrument. Students only report their experiences with discrepancies in learning and the feelings resulting from encountering such discrepancies.

Since iCARE was used on a regular basis (once per week), we observed in our class interactions that after a few times students start to get used to thinking about their learning processes during learning and become more fluent in responding to iCARE. This observation suggests that repeated use of iCARE can help train students to think more on the metacognitive side of their learning, which in turn improves the quality of students' responses to iCARE.

There are many sources that can lead to a student experiencing anxiety (e.g. test and grade-related anxiety). To ensure that the anxiety measured by iCARE is caused by cognitive conflict only, the entire iCARE instrument is built around students' experiences of discrepancies in learning and their feelings in responding to the discrepancies. The issue of grade-related learning achievement is not involved in the measurement. For example, if one were to measure grade-related anxiety, an item such as "I am concerned about my grades" would be appropriate. In iCARE, the items do not use any grade-related terms. In addition, with the style of the PBI course, students' grades are distributed among many participation-based activities. For example, attendance and participation in group discussion (regardless of whether correct arguments are given) make up about 20% of the student's grade. During a group learning session, students are engaged in group activities that involve many alternative conceptions, hypotheses, and discrepancies; therefore, encountering cognitive conflict is frequent in PBI classes and will not cue students into thinking about their grades. The iCARE instrument is available as supplementary material accompanying the online article.

# Validity and Reliability

iCARE builds on items in CCLT, which showed a content validity coefficient of 0.93 and a reliability coefficient between .69 and .86 (Lee et al., 2003). Therefore, iCARE inherits a significant portion of the validity and reliability of CCLT. Additional studies were conducted to further evaluate the validity and reliability of iCARE pertaining to the structure of the survey and alternations of the items from the original CCLT forms.

## *Content Validity*

The content validity of iCARE was assessed by 8 experts: 3 professors and 5 physics graduate students pursuing Ph.D. degrees in the Physics Education Research Program at OSU. All experts are very familiar with literature on cognitive conflict and anxiety and with the original CCLT instrument. In the context of iCARE, the graduate students are qualified as experts. They are part of the instruction team and have extensive teaching experience. They are well-versed in the ideas targeted by iCARE. Additionally, there is a very large content knowledge difference between the students being tested and the graduate student experts.

The experts used a 5-stage Likert scale to judge the validity of each item of iCARE. The judgment is based on two factors: (1) if an iCARE item is a close reproduction of a CCLT item, the expert will give a rating in 5 levels from 0 to 4 on if the iCARE item measures the same construct of the CCLT item, and (2) if an iCARE item is an extended alternation of the CCLT measures, the expert will give a rating from 0 to 4 on if the iCARE item measures a construct consistent with the consensus understanding from the literature on cognitive conflict and anxiety. A rating of 4 means very consistent whereas a rating of 0 means very inconsistent. The average rating of all items for each rater was then calculated, which ranged from 2.92 (0.73) to 3.44 (0.86) with a mean value of 3.24 (0.81). This validity rating is slightly lower than that of CCLT, which could be the result of the fact that iCARE measures a larger set of aspects related to cognitive conflict and anxiety and therefore involves a higher degree of freedom. When considering the added degrees of freedom, the rating of content validity of iCARE is quite comparable with that of CCLT.

### Response Validity

Validity based on response processes is focused on an analysis of responses to specific tasks and whether these responses are consistent with what is intended to be measured (McMillan & Schumacher, 2001). In order to identify evidence based on response processes, we conducted 13 interviews from which we analyzed the consistency between students' responses and explanations and what was intended to be measured. A high level of consistency would indicate that the items are well designed and can provide valid assessment on the intended measurement constructs.

The interview subjects were solicited with a small cash incentive from a pool of college students who were taking the PBI course at a large Midwestern University. Students were interviewed immediately after they finished their first class section so they had not had any prior exposure to iCARE. In the interviews, students were asked to take iCARE in one pass. Then students were asked explain aloud why they selected their answers. Students' answers and their explanations were recorded and transcribed. Students' explanations were compared with their answers to iCARE and rated with a score from 0 to 4 for consistency between their answers and their explanations (see Table 2 for the rating scheme).

Table 2.  The evaluation rubric for consistency between students' responses and their reasoning

| Score | Categories | Coding Criteria |
|-------|------------|-----------------|
| 0 | No explanation | The student replied "I don't know" or gave irrelevant explanation. |
| 1 | Vague explanation | The student attempted to explain but gave unclear explanations (little details) about what had happened in class that made him/her pick the answers.  The student also had little confidence about the explanation. |
| 2 | Partial explanation | The students' statements contain some but fuzzy details to what had happened in class that might have caused the conflicts.  The student had moderate confidence about the explanation. |
| 3 | Sound explanation | The student's explanation contains explicit details about experiment results or other group members' opinions that might have caused cognitive conflicts.  The student had moderately strong confidence about the explanation. |
| 4 | Very sound explanation | The student's explanation contains rich details about experiment results or other group members' opinions that caused cognitive conflicts.  The student had strong confidence about the explanation. |

Each student's interview data were evaluated by two researchers independently. For each item, the student responses and explanations were rated by the two researchers, who each gave a consistency score ranging from 0 to 4. If the two scores were within 1 level of difference, the average of the two was used as the consistency score. When the difference was larger than 1 level, a third researcher was invited to evaluate the data and give a rating score. The average of the closest two ratings was then used as the consistency score. From the interview results, the average consistency score of the 13 students was calculated to be 3.86 out of 4.0 with a standard deviation of 0.43. This result suggests that iCARE has validity with respect to students' response processes.  Note that the response validity has only been evaluated with college students; the applicability of this instrument to students in grade schools is unknown.

To determine whether the rating scheme is reliable or not, the inter-rater reliability was evaluated by measuring the consistency between the scores given by the two raters. In general, the consistency between the two raters is high. Although a third researcher was planned for resolving possible discrepancies between the two initial raters, he was not needed in the rating process. The two initial raters assigned identical ratings 92% of the times.

A statistical approach to determine inter-rater reliability is the Cohen's Kappa (Cohen, 1960), which is given in Eq. (1):

$$\kappa = \frac{P_O - P_R}{1 - P_R},$$ (1)

where $P_O$ is the observed percentage agreement and $P_R$ is the probability of random (chance) agreement. Kappa takes into account chance agreement by a renormalization using $(1-P_R)$ to remove the chance agreement probability. In this study, we have $P_O = 0.92$ and $P_R = 0.80$, which yielded a Kappa coefficient of 0.61. Usually, Kappa is characterized with a value over 0.75 as excellent, 0.40 to 0.75 as fair to good, and below 0.40 as poor (Fleiss, 1971; Landis & Koch, 1977). The result of this study is in the region of good agreement.

From the interviews, we also measured the time that a student takes to finish iCARE in one pass. We found that the average time was 8 minutes and 53 seconds with a standard deviation of 2 minutes and 46 seconds. Note that this was the first time the students ever took iCARE. According to our experience with about 120 students in three different classes, the average time the students took to finish iCARE dropped significantly to the level of 3 to 5 minutes for subsequent uses. This suggests that it is realistic time-wise to implement iCARE regularly in teaching.

Usually, to further establish the validity and reliability of an instrument, statistical tools such as factor analysis and measures of internal consistency are often used. However, iCARE measures a wide range of descriptive information that can be relatively open-ended when compared to other closed form surveys. For example, each learning session usually involves a dozen or so experiments that can each trigger a conflict or non-conflict situation for different students. Consequently, student reports on the contexts for cognitive conflict can involve a wide variety of combinations of these experiments. In addition, student experiences with these experiments were also quite diverse. As a result, the descriptive style of many iCARE items doesn't allow simple quantifiable scores to be extracted for correlational and consistency analysis. Therefore, factor analysis and internal consistency evaluation cannot be applied on the entire set of the iCARE items. However, with the cluster of items in part 2 of iCARE, we do have a case in which we can use factor analysis to study the construct validity.

There are three items in part 2. The first item measures the situation if a student is confused by the discrepancy occurred in learning. Items two and three both measure the appearance of a negative feeling in terms of discomfort caused by the confusion. By design, there should be two factors in this group of items. Factor 1 should load heavily on item 1 while factor 2 should load mostly on items 2 and 3. To verify if the iCARE items provide valid measurement on the two intended constructs, we performed a confirmation factor analysis on data collected from a 2004 class that had an enrollment of 32 students.

A correlation matrix from all students' responses to these three items accumulated throughout the entire quarter was calculated (N=245, p<0.000). The result is shown in Table 3. From the correlation matrix, it is obvious that the correlations between item 1 and either of items 2 and 3 are significantly lower than the correlation between item 2 and 3. The initial principal component analysis showed that the first two eigenvalues are much larger than the third one and the two together cover 91.6% of the total variance (see Table 3). This result confirms that there are two main factors. The rotated factor components were then calculated and also included in Table 3. We can see that factor 2 loads heavily on item 1 only while factor 1 loads equally heavily on both items 2 and 3 but not on item 1. These results confirm that the items in this cluster are valid in measuring two constructs that lead to an evaluation of student anxiety.

Table 3. Factor analysis results of iCARE Part 2 items

| Analysis | | Results | | |
|---|---|---|---|---|
| Correlation Matrix (N=245, 1-tailed, p<0.000) | | Item 1 | Item 2 | Item 3 |
| | Item 1 | 1.000 | 0.314 | 0.254 |
| | Item 2 | 0.314 | 1.000 | 0.744 |
| | Item 3 | 0.254 | 0.744 | 1.000 |
| Principal Components | Component | Initial Eigenvalues | | |
| | | Total | % Variance | |
| | 1 | 1.920 | 63.996 | |
| | 2 | 0.827 | 27.558 | |
| | 3 | 0.253 | 8.446 | |
| Rotated Component Factor Loading | | Component | | |
| | | 1 | 2 | |
| | Item 1 | 0.153 | 0.988 | |
| | Item 2 | 0.913 | 0.189 | |
| | Item 3 | 0.932 | 0.102 | |

In summary, we have conducted qualitative and quantitative research to develop, test, and refine iCARE. The results provide an initial evaluation of the validity and reliability of the instrument. However, establishing the validity and reliability of an instrument is always an ongoing process that cannot be treated in a once-and-for-all sense (AERA, 1999). The results discussed here are a starting point for further research on validating and refining iCARE.

### RESULTS: Application of iCARE in Research and Teaching

In the following sections of the paper, we report in detail a case study that applies iCARE in a PBI class at a large Midwestern university in 2004. This 10-week course covers topics on electric circuits.  On average, students work through one section per week. Thirty-two students were enrolled in this class. All students were second and third year non-science majors with half of them from the education department. Note that in some of the subsequent analysis, data from a smaller number out of the total thirty two students are used because some students missed certain parts of the class resulting in their data being incomplete and removed from that part of analysis.

The PBI course is a group-learning environment that implements an elicit-confront-resolve method of learning. Therefore, there are many situations designed to trigger cognitive conflict in students. Accompanying the inquiry method is a system of formative assessment and feedback through checkpoints, questions of the day, pretests, homework, exams, and journal entries.  iCARE was given to students as a post-evaluation to each section (except for Section 9, which was not evaluated due to scheduling issues). We asked students to complete the evaluation in class immediately after they finished work on the section. Having students complete the evaluation right after each section was intended to improve the quality of the data; however, the evaluation is still a "self-reporting" method and is subject to the drawbacks of this type of method.  Part of Table 4 lists the sections in which we used iCARE as the post-evaluation.

The main goal of this study is to demonstrate the type of information that can be obtained by using iCARE and how data from iCARE can be analyzed for various purposes in teaching and research. Due to the small class setting of the PBI course, the results of this study may not support strong comparative claims;

however, the outcome is useful in making indicative inferences for researchers and instructors when using iCARE in research and teaching.

In an inquiry-based course that uses cognitive conflict as a constructive teaching strategy, there are several research questions that often attract the attention of teachers and researchers:

(1) What curriculum components and contexts are likely (or unlikely) to trigger cognitive conflict?

(2) What types and fractions of students are likely (or unlikely) to experience cognitive conflict?

(3) Which education settings, such as hands-on experiment or group discussion, are more or less likely to trigger cognitive conflict?

(4) How do students affectively react to experiencing cognitive conflict and how do such reactions vary with students of different backgrounds and at different levels of learning.

(5) What levels of anxiety do students experience in dealing with cognitive conflict? How does anxiety change over content areas and with different students?

(6) For students with high or low anxiety, how do they react in their learning behaviors to address cognitive conflict?

(7) With all the questions above, how do the concerned factors vary with the type of student, time, and progression of learning?

In the following sections, we will demonstrate how information collected with iCARE can answer the questions listed above.

### Course Components and Students' Cognitive Conflicts

The data discussed in this section can help answer parts of the first and second research questions. In this study, the data collected with iCARE reveal that more than 90% of the students reported at least one conflict in each topic section (see Table 4). We have not yet found in the literature previous reports on the quantitative frequencies of student experiences with cognitive conflict in an inquiry-based course, so this result is an important base-line data point that can help set the context for both teaching and further research in this area. The high concurrency of cognitive conflict is consistent with the defining feature of PBI, which, by design, intentionally triggers students' conflicting ideas and guides them to resolve the conflicts through a series of inquiry learning tasks.

Table 4. Numbers of reported cognitive conflicts in different sections

| Section | Topic | Number of Exercises/ Experiments in a Section | Number of Students Reporting Occurrences of Conflicts in a Section* | | | | | | N** |
|---------|-------|------------------------------------------------|-----|-----|-----|-----|-----|-----|------|
| | | | 0 | 1 | 2 | 3 | 4 | 5 | |
| 1 | Single-bulb circuits | 14 | 3 | 4 | 4 | 1 | 19 | 0 | 31 |
| 2 | Model for current | 7 | 1 | 5 | 4 | 2 | 20 | 0 | 32 |
| 3 | Extending model for current | 10 | 2 | 5 | 4 | 0 | 13 | 8 | 32 |
| 4 | Series/parallel networks | 14 | 2 | 5 | 1 | 2 | 1 | 18 | 29 |
| 5 | Kirchhoff's first rule | 9 | 2 | 5 | 1 | 1 | 0 | 14 | 23 |
| 6 | Equivalent resistance | 5 | 2 | 5 | 1 | 4 | 1 | 11 | 24 |
| 7 | Multiple batteries | 16 | 2 | 5 | 3 | 0 | 0 | 13 | 23 |
| 8 | Kirchhoff's second rule | 18 | 1 | 7 | 1 | 0 | 0 | 12 | 21 |
| 10 | Ohm's Law | 14 | 1 | 8 | 1 | 0 | 1 | 19 | 30 |
| | Total | 107 | 16 | 49 | 20 | 10 | 55 | 91 | 245 |

\*   0 means no conflicts; 5 is the maximum number of conflicts that a student can report.
\*\* N is the number of students who completed iCARE for the given section.

From Table 4, we can see that the number of students who reported having cognitive conflict is relatively consistent throughout the 10-week course period. This result suggests that the PBI curriculum offers a reliable education structure that consistently employs the cognitive conflict oriented inquiry learning. The result also shows that students respond well to iCARE, which further establishes its validity in engaging students in the measurement.

A closer look at the data further reveals that in most topic sections about one-third of the students experience zero to two conflicts while the remaining students experience three or more conflicts. It appears that we can categorize students into two groups: a low-conflict group and a high-conflict group. Since very few students reported experiencing "3" cognitive conflicts, it suggests that the number "3" works well as a threshold to separate the low- and high-conflict groups.

A detailed analysis of the individual students' data confirms the observation: out of the total 32 students, about one third often consistently experience a low number of cognitive conflicts throughout the course, while the remaining two thirds of the students consistently experience a higher number of cognitive conflicts. Only a few students (≤5) moved between the low- and high-conflict groups at different topic sections.

Another interesting observation is that among the students in the high-conflict group, there is a clear shift from reporting "4" conflicts to "5 and more" conflicts at topic section 4, which starts to teach more advanced circuit manipulations. This shift is consistent with the expert view of the difficulty levels of the content topics. Since the shift only occurs in the high-conflict group, the result suggests that the level of content difficulty didn't impact the learning experiences of the low-conflict group students.

The results and analysis show that iCARE can probe quantitative information about how different students respond to components of the instruction regarding the activation of cognitive conflict. This kind of data can provide useful evidence for a range of studies such as learning behaviors of students with different backgrounds and efficacy of curriculum in using cognitive conflicts for constructive inquiry.

### Education Contexts Triggering Cognitive Conflicts

This part of analysis responds to the third research question listed in the beginning of this section. The design of iCARE allows us to probe the four categories of education contexts discussed in the design section. These are coded as Experiment, Concept, Group, and Teacher. Experiment represents conflict induced by differences between students' conceptions and observations in an experiment. Concept represents conflict induced by differences between a student's different conceptions. Group represents conflict induced by differences among a student and his/her group members. Teacher represents conflict induced by differences between a student and the teacher.

The data show that the education context situations reported as leading to cognitive conflict vary from student to student and over different content topics. Figure 2 shows the relative percentages of the different situations causing cognitive conflict out of the total number of cognitive conflicts reported in each section. We can see that, on average, about 50% of the conflicts are the Experiment type. Note that the Experiment type of conflict is often the focus of many researchers (Hashweh, 1986; Hewson & Hewson, 1984; Piaget, 1950) and represents a commonly used method in curriculum development (McDermott et al., 1996). An interesting result is that among the four categories of conflict, the Teacher type is much less frequent than others, showing that in the PBI course cognitive conflict is mostly induced by students' interactions with experiments and group members and with their own alternative conceptions rather than with instructors. These results are consistent with the context and instructional goals of PBI, which are learner-centered and emphasize more active roles for the learners rather than the instructors.

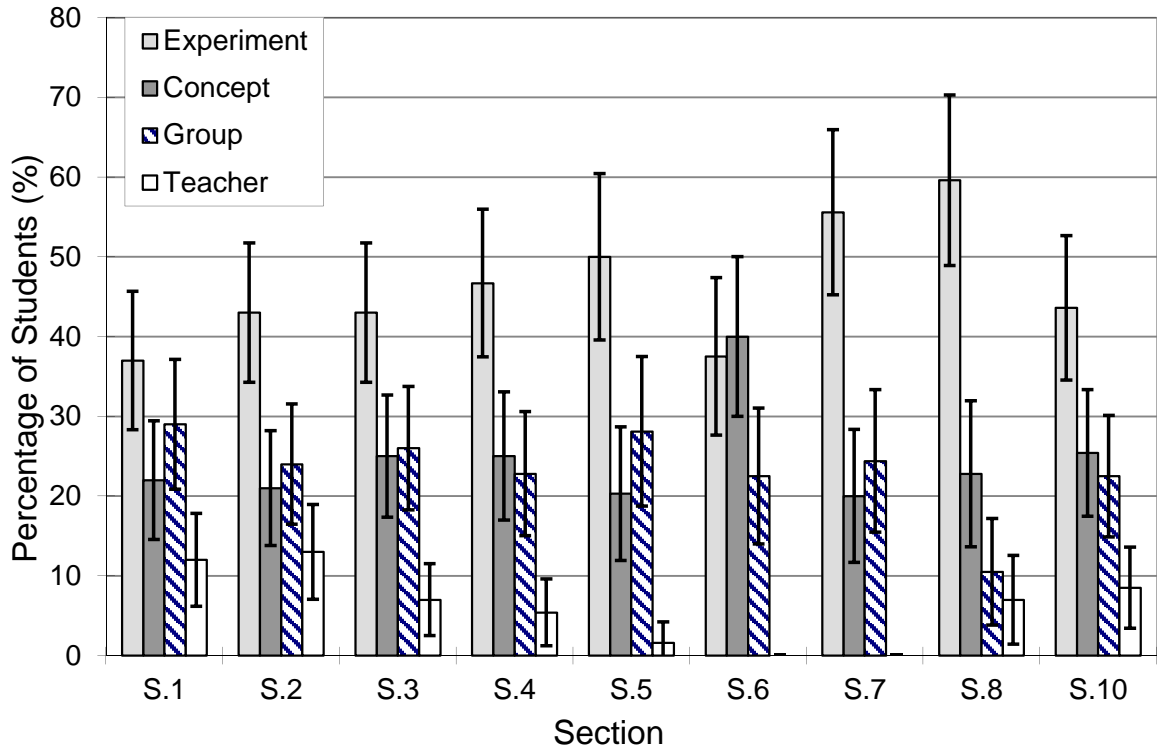

Figure 2.  Relative proportions of different situations causing cognitive conflicts. The error bars reflect the standard errors of the results.

From Figure 2, we can also see that the conflict contexts vary over different content topics. For example, the Concept type has the largest percentage (40%) in topic section 6 compared to other topic areas (average about 23%). In this topic section, students learn a new method to find equivalent resistance by combining several previously learned concepts on resistance. During the learning, students often develop multiple alternative understandings on how to determine the equivalent resistance. The education settings use a series of prediction tasks that require students to first make a prediction of the equivalent resistance of a given circuit and then check their predictions with experiments and group members. The prediction tasks provide outlets for a wide variety of the students' own conceptions; therefore, it is not surprising that students reported a higher percentage of conflicts coming from discrepancies between different understandings.

The variations in conflict type over content areas and the possible inferences of the data suggest that iCARE is sensitive to curriculum constructs and can be used to evaluate the effects of different aspects of curriculum on triggering cognitive conflict. The measured information about the different education contexts is particularly valuable for evaluating the effectiveness of curriculum and instruction in delivering inquiry-based learning. Such information can also help instructors and researchers revise instruction in order to promote certain types of cognitive conflict in learning for specific teaching goals.

### Students' Affective Reactions to Conflict Situations

When cognitive conflict is encountered, students may experience different feelings. Based on the measures in CCLT, iCARE includes three items to probe such feelings: "surprised", "interested", and "trying to pay more attention to the topic." The measurement constructs underlying these items combine two measurement components from CCLT: "recognition of contradiction" and "interest." These components form the measure of students' recognition of experiencing a cognitive conflict (Lee et al., 2003). The results of this part of measurement can help answer the fourth research question listed in the beginning of this section. The benefit of measuring these affective feelings is not only that they provide a measure of student recognition of cognitive conflict but also that these feelings, which are generated in responding to anomalous situations, can affect future steps of student learning, especially the ways students may address the discrepancies. Research has shown that students may try to reconcile the cognitive conflict when enough interest and attention are triggered; however, students may also ignore the conflict without proper cognitive and affective motivation (Strike & Posner, 1992, Pintrich, Marx, & Boyle, 1993).

More generally, many researchers have emphasized that the process of conceptual change is affected by a range of affective variables and value beliefs (Strike & Posner, 1992; Pintrich, Marx, & Boyle, 1993; Pintrich & Sinatra, 2003; Limón, 2001; Kim & Kwon, 2004). When a cognitive conflict is encountered during learning, in order for students to start a process of conceptual change, the cognitive conflict has to be "meaningful" for the student, which means that the student must be motivated and interested in the topic, activate their prior knowledge, and have adequate epistemological beliefs and reasoning abilities to deal with the given problem. In particular, students' personal interest in a topic might determine whether they even attend to a discrepancy that could lead them to dissatisfaction with their existing conceptual understanding. Therefore, the data collected on students' feelings will provide information about students' affective status during their conceptual change process. This kind of information can be very useful for researchers and teachers in understanding how students' conceptual changes are affected by factors in the instruction context and in evaluating the extent to which a particular curriculum component promotes favorable conceptual change.

The results of this part of measurement are shown in Figure 3, which gives the relative proportions of reported feelings out of the total number of cases reported in each section. The design of iCARE allows students to report more than one type of feeling; however, the data shows that fewer than 5% of students reported multiple feelings. The results in Figure 3 are calculated with all reported cases; the effects of multiple reports by a single student in this study are small and are not analyzed separately.

From Figure 3, we can see quite a range of variations among the different reported feelings. Students can encounter many surprises in certain topic areas (e.g., S2), while other topic sections can generate a lot of interest (e.g., S1 and S5). Students can also encounter less interest and surprise but pay high attention to the content topics (e.g. S7 to S10). These variations reveal unique and interesting information about the possible interactions between students and the teaching materials and methods, which seems to have not been extensively addressed in the existing literature. Such information is useful for understanding students' learning processes in real education settings and for studying the possible causal factors of curriculum components that impact learning.

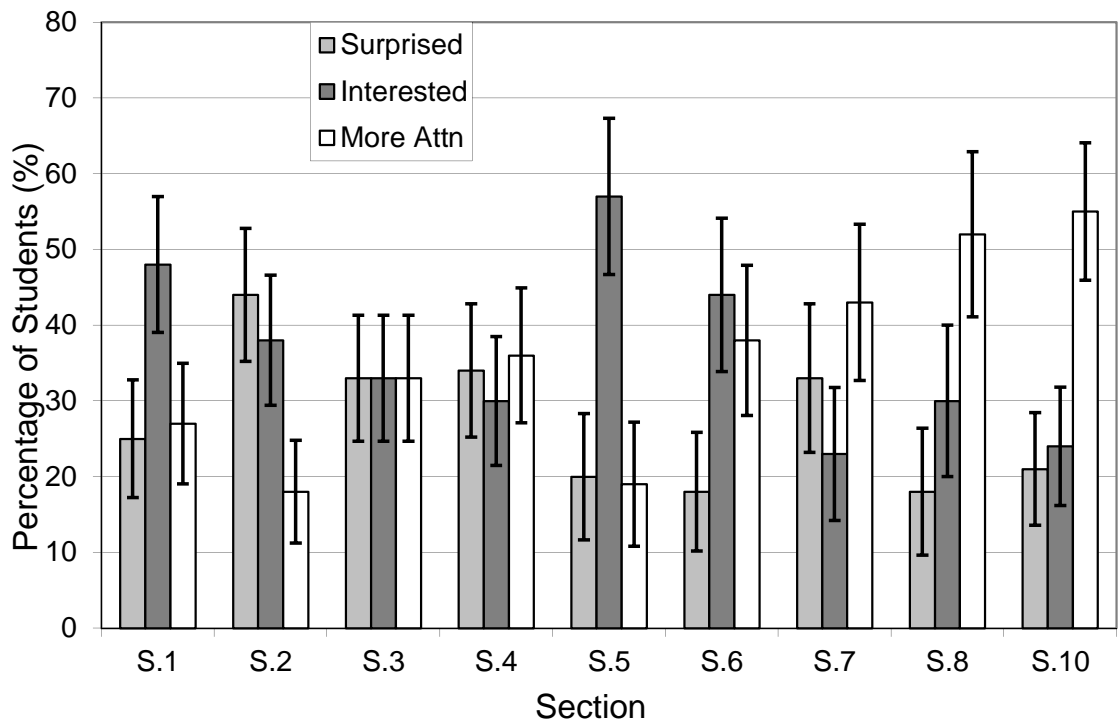

Figure 3.  Relative proportions of students' reported feelings. The error bars reflect the standard errors of the results.

For example, with the data shown in Figure 3 we can make several inferences that may warrant detailed future studies. One is that among the nine topic sections, S1 and S5 show a significantly higher level of interest among students. This may suggest that the topics or education methods in these two sections can have unique features generating more interests to students. Comparisons of these topics and others can produce further insight on what may trigger more or less interest among students.

Another interesting observation is as the course moves towards the end of the quarter, the number of conflicts reported remains stable (see Table 4), but the patterns of reported feelings change rather obviously. The number of students reporting "interested" decreases from S5 to S10, while the number of students reporting "paying more attention" increases.  This pattern may be caused by two factors. One is the increasing difficulty of topics towards the end of the course and the other is exam-related. Classroom discussions with students suggest that students became more concerned about their correct understanding of the content as the course approaches the final exam and thus paid more attention to the topics leading to cognitive discrepancies. The results also suggest that students' interest and attention can be as dependent as they are independent. For example, it is not uncommon in a college physics course for students to be uninterested by a content topic but yet pay serious attention to it if it will be included in the exam. This is not to say that the experiences of conflict and anxiety depend on time (these actually depend more on content, as discussed in the next section), but rather that students may react to conflict in different ways depending on the different stages and contexts of the course.

From the examples discussed above, we can see that using iCARE can produce various types of new data supporting studies to address core research questions about the impacts of both cognitive and affective factors on the interactions between students and education settings.

***Students' Cognitive Anxiety and Learning over Content Topics and Student Performance***

In this part of discussion, we explore how iCARE can be used to answer the fifth and sixth research questions concerning students' cognitive based anxiety and their learning. The third part of iCARE gives a numerical estimate of a student's anxiety level. Students rate, using a 5-point Likert scale, if they are confused, uncomfortable, or upset about the cognitive conflict they have encountered. These affective responses are typical feelings that signal the experience of anxiety, and the ratings of these are then summed to produce an anxiety score (Lee et al., 2003).

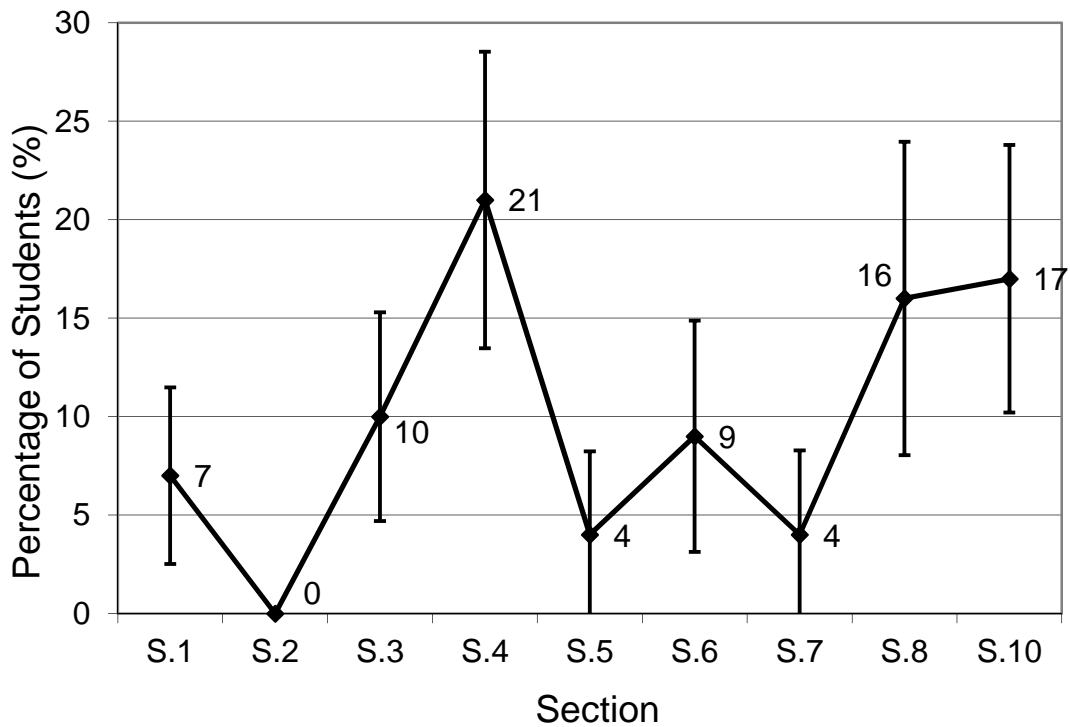

Figure 4.  Students revealing a high level of anxiety in different sections. The error bars reflect the standard errors of the results.

The first question we explore is if students' anxiety is affected by content topics. Students' anxiety scores, which range from 3 to 15, are coded into two levels. The low level includes anxiety scores from 3 to 8, and the high level is for anxiety scores from 9 to 15. Figure 4 shows the percentage of students reporting a high level of anxiety in the different topic sections. The results suggest that in general only a small fraction of students experience high levels of anxiety during learning, and that the anxiety level also varies with content topics. For example, Section 4 triggers the highest fraction of students revealing a high level of anxiety. When contrasting these results with the data shown in Table 1, we can see that Section 4 is also the starting point when many students begin to report 5 or more conflicts. Therefore, the change of difficulty level in content seems to have also impacted student anxiety. Although about two thirds of the students typically reported 4 or more conflicts in a topic section, much fewer (20% or lower) experienced a high level of anxiety. This indicates that many students were able to resolve the conflicts without a high level of affective stress. As shown by research, anxiety is not necessarily negative to learning; rather, a small amount of anxiety may facilitate learning, especially if the task is not too difficult (Ball, 1995). Based on the results of this study, one can further infer that the curriculum design and education method of PBI is appropriate in helping students learn by constructively addressing their conflicts and revising their alternative conceptions.

The second question we explore is if student anxiety and course performance are related. Due to the small sample size and the small percentage of students revealing a high level of anxiety, we choose to focus this part of analysis on data from Section 4 in order to have the largest possible group of students with high anxiety in learning. In addition, a midterm exam was given to students one week after they finished Section 4. This midterm is largely focused on the topics in Section 4; therefore, there is a pedagogical connection between the Section 4 content and the midterm assessment, which provides a context for possible links between learning and exam performance.

For this part of the study, students are divided into two groups based on their anxiety scores measured in Section 4 of the course. The low anxiety group contains twenty-three students while the high anxiety group contains six students. The class contains a total of 32 students, out of whom three were missing part of the class and were removed from this analysis. Out of the maximum of 15 points on the anxiety score, the low anxiety group has an average score of 5 points while the high anxiety group has an average score of 10 points.

A comparison of the midterm scores of students with different anxiety levels is included in Table 5. The results show that the students who experienced high anxiety in Section 4 also had lower midterm exam scores. This may suggest that students with weaker academic background often experience a higher level of anxiety during learning. Another possibility is that failure to productively resolve conflict can negatively impact students' learning achievement and induce higher anxiety. Therefore, how students react to cognitive conflict under the influence of anxiety can play an important role that determines whether the students will be able to achieve favorable learning outcomes.

Table 5. Comparison of first midterm exam scores of students with different levels of anxiety in Section 4

| Anxiety Level | N | Exam Mean | SD | t | p | Effect Size |
|---------------|----|-----------|------|-----|------|-------------|
| Low | 23 | 89% | 6.0% | 2.4 | 0.02 | 0.9 |
| High | 6 | 81% | 10% | | | |

As suggested in previous research, students of different backgrounds can react differently in learning with respect to the impacts of anxiety (Cassady & Johnson, 2002; Hadar & Hadass, 1990). This leads to our third question about anxiety regarding how students with high anxiety may react to anxiety and cognitive conflicts.

The fourth part of iCARE measures in more detail students' affective reactions to their cognitive conflict under the affective influence of the induced anxiety. The reactions were coded into the eight categories shown in Table 1. From this study we found that among the six students exhibiting high anxiety in Section 4, two students exhibited "Re-inspection of reasoning for predictions." Two students exhibited both "Recognition of inability to resolve conflict" and "Re-inspection of reasoning for predictions". One student exhibited "Lack of self-confidence" and "Confidence in preconceptions." The remaining student had "Re-inspection of reasoning for predictions" and "Confidence in preconceptions." It appears that most high anxiety students had a tendency to hold on to their existing understanding and decide that the conflict was produced by inappropriate application of their understanding (rather than recognizing that their understanding was unproductive). These students didn't see immediately the need for changing their understanding, and therefore they were often unable (or unwilling) to revise their preconceptions to the expert views. This kind of information and possible inferences are important for understanding the conceptual change processes of students who are encountering learning difficulties. Research into this area can also shed light on instructional strategies that may help students constructively resolve conflicts and control their anxiety levels.

In this study, instructors followed typical PBI instruction methods. They did not implement any additional strategies to address the anxiety issue. Educational psychologists have developed methods for alleviating test anxiety (Schutz & Davis, 2000; Zeidner, 1998), but these are not appropriate for use in addressing anxiety caused by cognitive conflict in a PBI class. Adapting anxiety control in inquiry-based learning warrants new research, which can also be supported by iCARE.

Fully understanding the possible interactions between anxiety and learning will require additional substantive studies deep into the process of learning. iCARE provides a new tool to facilitate such studies. For example, iCARE can help identify what kind of anxiety or how many anxiety points in a given situation assist rather than hinder learning. The results discussed here show an example of using iCARE in research that address the cross-connections among affective variables, context variables, and learning performance.

### *A Case Study of Possible Interactions between Curriculum Components and Cognitive Conflicts*

In this section, we present a sample case study to investigate at a small grain size how specific pieces of curriculum may affect students' cognitive conflicts. In particular, we explore two research questions concerning students' cognitive understanding, learning behavior, and test performance.

The first question we explore is whether there is a connection between students' cognitive understanding (such as alternative conceptions) and their learning behaviors in terms of cognitive conflict.

We choose to study student learning in the content topic of Section 4, in which the largest fraction of students had shown a high level of anxiety. Table 6 lists all the learning activities in Section 4 and the number of students who reported experiencing cognitive conflicts within each of the activities.

As shown in Table 6, Section 4 has 14 learning activities (7 exercises and 7 experiments) on series and parallel circuits. On average, about one-quarter of the students (27%, SD=12%) reported cognitive conflicts in these tasks. The number of students reporting conflicts also varies with the complexity of the content. For example, many students reported conflicts in Exercise 4, which is the first in the list of activities of Section 4.

Table 6. Students experiencing cognitive conflicts in different learning activities in Section 4

| Learning Activities | Content Topics | Students Reporting Conflicts (N = 29) | |
|---|---|---|---|
| Exercise 1 | Define series and parallel connections. | 3 | 10% |
| Exercise 2 | Apply definitions of series and parallel connections. | 4 | 14% |
| Exercise 3 | Identify series and parallel connections and networks. | 5 | 17% |
| Exercise 4 | Rank circuits in terms of the current through the battery. | 10 | 35% |
| Experiment 5 | Predict the brightness of bulbs in circuits. | 11 | 38% |
| Experiment 6 | Compare two students' comments about the brightness of bulbs in circuits. | 2 | 7% |
| Experiment 7 | Predict and observe the change in brightness when the switch in a circuit is opened and closed. | 12 | 41% |
| Experiment 8 | Predict and observe the brightness of the bulbs in more complicated circuits. | 11 | 38% |
| Exercise 9 | Categorize series and parallel circuits. | 4 | 14% |
| Exercise 10 | Match freeform and standard circuit diagrams. | 7 | 24% |
| Exercise 11 | Match freeform and standard circuit diagrams (harder). | 9 | 31% |
| Experiment 12 | Analyze functions of a circuit with SPDT switches. | 13 | 44% |
| Experiment 13 | Design room lights control using SPDT switches. | 11 | 38% |
| Experiment 14 | Analyze functions of a given circuit board. | 6 | 21% |

For a more substantive analysis, the questions used in Exercise 4 are shown in Figure 5. In these questions, students were asked to redraw circuit diagrams and to rank the circuits according to the current through the battery. Problems similar to the ones in Exercise 4 have been used in many studies on student learning of circuits (McDermott & Shaffer, 1992; Shaffer & McDermott, 1992; Engelhardt & Beichner, 2004), which have shown that students often hold strong beliefs about the battery being a source of fixed current. This type of understanding is also related to difficulties in Experiments 5, 7, and 8, in which students were asked to analyze the brightness or change in brightness of bulbs in various circuits. In order to solve these problems, it is crucial for students to understand that an ideal battery supplies a different current depending on the external circuit (i.e., a battery is not a fixed current source). These are quite challenging tasks and students often make incorrect predictions based on their misconception about a battery being a constant current source.

Exercise 4.
    Suppose you have three boxes, labeled A, B, and C. Each box has two terminals. The arrangement of bulbs inside each box is shown below.

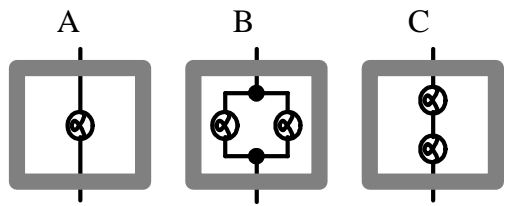

A. For each of the following circuit, draw a standard circuit diagram showing all the bulbs in the circuit. List the series and parallel combinations for each circuit.

B. Rank each of the circuits in part A according to the current through the battery.

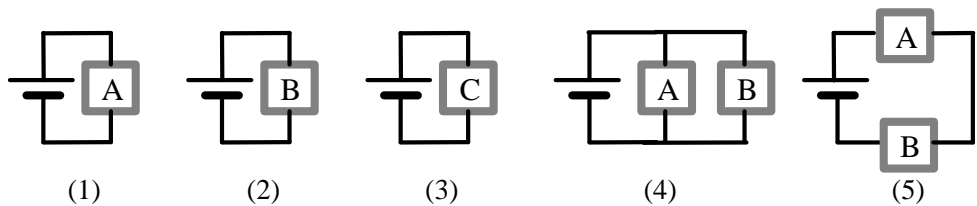

Figure 5. Problems used in exercise 4 of Section 4 of the PBI curriculum. (recreated from Exercise 4.4 of Physics by Inquiry).

From Table 6, we can see that in all three experiments (5, 7, and 8) a relatively large number of students have reported conflicts. This result is consistent with the cognitive research about student misconceptions in this area. The agreement between the cognitive studies and iCARE measures can further establish the validity of the iCARE instrument. It also shows that iCARE can provide a new venue of measures that connect the cognitive traits of students' conceptions to their actual learning behaviors in real education settings.

The second question we explore is whether there is a connection between students' learning behaviors (in terms of cognitive conflict and reactions to conflict) and their test performance. In our studied course, one week after students finished learning Section 4, a midterm exam was given, which included a question similar to the one used in Experiment 7. This question is then used as the basis to compare students' test performance with their learning behaviors in the classroom. The problems used in Experiment 7 and on the first midterm are shown in Figure A1 and A2 included in the supplemental materials.

Table 7. Students' performance on the midterm question

| Students reporting conflicts in Section 4 (N=29) | Answered Correctly | Answered Incorrectly |
|---|---|---|
| No conflicts in entire Section 4 | 2 | 0 |
| With conflicts in Experiment 7 | 8 | 4 |
| With conflicts, but not in Experiment 7 | 7 | 8 |

Among the twenty-nine students in the class (coded with S1 … S29), two reported no conflicts throughout Section 4; twelve reported having conflicts in Experiment 7; the remaining fifteen students reported having conflicts in activities other than Experiment 7. The students' performance on the midterm question (see Figure A2 in the Appendix) is given in Table 7.

Table 8. Reactions of students who reported conflicts in Experiment 7 and their performance on the midterm question

| Students who reported having conflicts in Experiment 7 | | | Students' performance on the midterm question | |
|---|---|---|---|---|
| Anxiety Level | Reactions | Students | Correct | Incorrect |
| Low (9 students) | Agreed predictions | S10, S11, S18, S29 | 8 | 1 (S29) |
| | Confidence in resolving conflicts | S9, S10, S11, S18, S19, S20, S22, S29 | | |
| | Dependence on others' ideas | S8 | | |
| | Use of past personal experience | S10, S18, S20, S29 | | |
| High (3 students) | Confidence in preconceptions | | 0 | 3 |
| | Re-inspection of reasoning for predictions | S1, S2, S12 | | |
| | Lack of self-confidence | | | |
| | Recognition of inability to resolve conflict | S1 | | |

Due to the small sample size, the results are interpreted as case studies and do not imply statistically significance. However, these can be used as a practical example showing how such analysis can be carried out at a larger scale to further identify possible relations between students' experiences of conflict and their test performance. For example, the data in Table 7 could imply a trend that students who reported having conflicts in Experiment 7 perform better than students who reported conflicts in experiments other than Experiment 7. A possible speculation based on this type of data could be that explicit recognition of conflict may help students' learning, which can lead to further studies on this observation. The results reported here serve as a demonstration of how iCARE can be used to carry out such research.

Students' performance on the test question and their reactions to cognitive conflict are also compared. Among the twelve students who reported having conflicts in Experiment 7, three had a high level of anxiety and nine had a low level of anxiety. Of the nine students with low anxiety, eight answered the midterm question correctly. All three students who exhibited high anxiety answered the midterm question incorrectly (see Table 8). This result isn't surprising since the students who had low anxiety were often higher achieving students. What is useful is that by using iCARE, we can collect detailed information about students' reactions during the learning process, particularly for those who did poorly on the test. This will allow us to build an understanding of possible common behavior patterns of students who may be successful (or unsuccessful) in learning. Such information can also help instructors identify at-risk students while learning is taking place (not after the exam) so that proper treatment can be implemented during the course of instruction.

The analysis in this section shows that by combining the results from iCARE and detailed content analysis, researchers can gain insights into how specific curriculum components affect student learning in both cognitive and affective dimensions. This can help develop better instructional materials that would utilize the affective factors in creating a more effective learning environment.

### Possible Additional Uses of iCARE in Teaching and Research

iCARE can be used for several purposes other than as a measurement tool including formative assessment, student self-reflection on the process of learning, and curriculum development.

Awareness of student beliefs and attitudes can provide guidance to instruction (Limón, 2001). In a course such as PBI, iCARE can clue instructors in to which students are struggling and potentially being left behind by group members. In a group setting, it is not always easy for students to admit they are lost. iCARE provides an outlet for this frustration. Knowing that iCARE does not affect one's grade and is seen only by the instructor makes students more likely to submit their true feelings. Additionally, iCARE can guide instructors in asking appropriate questions that lead to productive discussions with students. The PBI course has checkpoints throughout the lessons where students interact for an extended period of time with the instructor. If the instructor knows the individuals who are experiencing unresolved conflicts, he/she can specifically target the troubled students with more care. On the other hand, for students who rarely feel anxiety, more challenging questions can be posed and such students can also be encouraged to help struggling group members.

Students can benefit from using iCARE as well. Students who are conscious about their learning will get more out of each classroom experience (Elby, 2001). Engaging students in frequent self-reflections about the learning processes (rather than always focusing on the content) is one important goal of iCARE. After a section of learning, students pause to reflect on what just happened during class and re-think about the entire process of their learning. This is rarely conducted in traditional instruction but is of great value to gradually help students develop a more conscious level of appropriate meta-cognition. iCARE gives students an opportunity to think about what they have done in learning from a different perspective that draws more explicit attention to the process of scientific inquiry.

While taking iCARE, students don't have to worry about the content; instead they focus on how they experience learning in an inquiry-based environment. In a sense, iCARE provides an opportunity for systematic epistemological education in which students are repeatedly reminded that it is okay to have cognitive conflicts and that such conflicts are necessary constructive steps in scientific inquiry.

In the context of education research, iCARE provides a new tool that can guide curriculum development. Using iCARE can help instructors identify curriculum units where students show high levels of anxiety. Points where students would benefit from instructor intervention can also be identified. Based on the information gathered from iCARE, curriculum can be optimized to avoid, for example, having several class periods in a row where students are likely to have high anxiety. In this way, iCARE serves as an aid in regulating and maintaining a "healthy" level of cognitive based anxiety to produce a more effective inquiry based learning environment.

## SUMMARY and IMPLICATIONS

Inquiry based teaching and learning methods that engage students in an active process of constructing knowledge are becoming the basis for current-day education development and practice. The emphasis on the process of learning also demands consideration and control of a new set of variables such as social affective factors that cannot be assessed using only performance-based measures (Pintrich & Sinatra, 2003; Snow, Corno, & Jackson, 1996). Therefore, it is important to develop assessment tools that probe behavior-related affective factors. This study is one such attempt. We developed an instrument (iCARE) to probe a number of affective factors related to the learning processes in commonly experienced education settings. The results shown in this paper are for the purpose of exemplifying the types of information that can be obtained with such an instrument and how such information may be used in further research and teaching practices.

From this study, we can foresee interesting possibilities for using iCARE in research and instruction. For example, researchers can use the instrument to inspect how specific curriculum components affect student learning in terms of triggering cognitive conflict and causing anxiety. From students' learning behaviors and reactions to conflict situations, one can further obtain additional assessment of students' preparation and learning styles. This instrument can also be integrated into a formative assessment framework to directly benefit students.

# Appendix

Questions used in Experiment 7 of Section 4 and on the midterm exam.

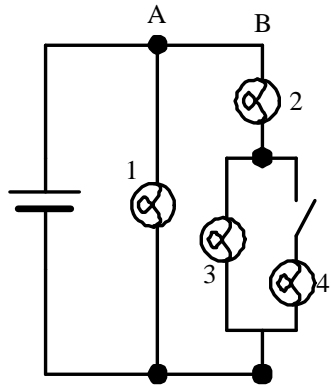

Experiment 7.

(A) Observe the change in brightness of bulb 2 when the switch is opened and closed. Now observe bulb 1 as the switch is opened and closed.

(B) Predict the effect on branch A (or B) of each of the following alterations of branch B (or A):
(1) Unscrewing bulb 2
(2) Shorting out bulb 3
(3) Unscrewing bulb 1
(4) Adding a bulb in series with bulb 1

Figure A1. Problems used in experiment 7 of Section 4 of the PBI curriculum. (recreated from Experiment 4.7 of *Physics by Inquiry*).

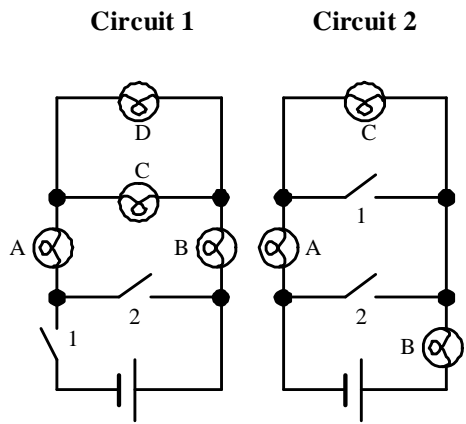

(P) Rank the bulbs by name from brightest (first) to dimmest (last) for each circuit (1 and 2), under the following condition
(1) Both switches are open;
(2) Switch 1 is open, switch 2 is closed;
(3) Switch 1 is closed, switch 2 is open;
(4) Both switches are closed.

Figure A2. The problems used in the first midterm exam given one week after Section 4.

# iCARE Instrument for Scientific Inquiry Labs

During the class, you may have encountered situations that caused:
  (**1**) Differences between your predictions (or what you believed) and the results of an experiment.
  (**2**) Differences between your understanding of one experiment and your understanding of another experiment.
  (**3**) Differences between your opinions and the opinions of other group members.
  (**4**) Differences between your opinions (or what you believed) and the opinions of the instructor.
  (**5**) Complete agreement with your opinions, instructor's opinions, and experiment results.

When someone encounters such differences, he/she may have different kinds of experiences such as
  A. *The differences surprised me.*
  B. *The differences increased my interests in the topic.*
  C. *The differences made me want to pay more attention to the topic and spend more time to work on it.*

In the following table, please identify the experiments that may have given rise to the different situations discussed above. Identify the situations with (1) ~ (5) and your experience with A, B, C (see above). **Select all that apply**. You may add your own categories if not listed. If you need more space and/or have more comments, use the back of the page.

| Experiment ID | The situation that caused differences | Your experiences with the difference | Now, can you completely resolve the difference by yourself? |
|---|---|---|---|
|  | ☐ (1)  ☐ (2)  ☐ (3)  ☐ (4)  ☐ (5) | ☐ A  ☐ B  ☐ C | ☐ Yes  ☐ No |
|  | ☐ (1)  ☐ (2)  ☐ (3)  ☐ (4)  ☐ (5) | ☐ A  ☐ B  ☐ C | ☐ Yes  ☐ No |
|  | ☐ (1)  ☐ (2)  ☐ (3)  ☐ (4)  ☐ (5) | ☐ A  ☐ B  ☐ C | ☐ Yes  ☐ No |
|  | ☐ (1)  ☐ (2)  ☐ (3)  ☐ (4)  ☐ (5) | ☐ A  ☐ B  ☐ C | ☐ Yes  ☐ No |
|  | ☐ (1)  ☐ (2)  ☐ (3)  ☐ (4)  ☐ (5) | ☐ A  ☐ B  ☐ C | ☐ Yes  ☐ No |

From the experiments you listed, select one that had the most impression to you and use it as the basis for answering the questions listed below: ▶ **Write down the experiment ID that you have selected (_____________).**

1. The result of this experiment confused me.
  1　　2　　3　　4　　5
  NOT AT ALL TRUE　SOMEWHAT TRUE　VERY TRUE

2. Since I cannot resolve the differences, I am uncomfortable.
  1　　2　　3　　4　　5
  NOT AT ALL TRUE　SOMEWHAT TRUE　VERY TRUE

3. I am upset because I cannot understand the reason for the result.
  1　　2　　3　　4　　5
  NOT AT ALL TRUE　SOMEWHAT TRUE　VERY TRUE

4. Sum up your answers to the above three questions and put **the total number** here: (_____________)

▶ If your number in 4 is less than 9 (**3~8**), go to **Part 1** only. If your number is **9~15** go to **Part 2** only.

▶**Part 1 (Finish this part if your calculated number is 3~8):** Among the following statements, **check the item** that best describes the likely causes of the feelings you reported above.
  ☐ 1. Before the experiment, I predicted multiple possible outcomes. From the experiment, I have seen one of my predictions proved. So I am satisfied with the experiment result even without detailed explanations.
  ☐ 2. I was confident that by reevaluating my previous beliefs, I would be able to find an explanation without others' help.
  ☐ 3. I accepted what instructors or my classmates had said. I didn't spend much effort to find an explanation on my own.
  ☐ 4. I made my predictions for this experiment by thinking about my past experience. I also tried to make sense of what I saw in the experiment based on my understandings of my past experience.
  ☐ 5. Others (please specify. Use the back of the page if necessary.)

▶**Part 2 (Finish this part if your calculated number is 9~15):** Among the following statements, **check the item** that best describes the likely causes of the feelings you reported above.
  ☐ 1. Before the experiment, I was highly confident in my original understandings of the subject. However, my understanding seems to be inconsistent with the outcome of the experiment.
  ☐ 2. After I saw the outcome of the experiment, I tried to explain it by considering things that I might have ignored when I was making my predictions.
  ☐ 3. I believe that there must be good reasons that can explain the experiment well. But right now I don't think I have learned enough physics to build a good explanation yet.
  ☐ 4. On this experiment, the results are inconsistent with what I expected based on my experience and I haven't been able to resolve the discrepancies yet.
  ☐ 5. Others (please specify. Use the back of the page if necessary.)